\documentstyle[psfig,jkas35]{article}

\runningauthor{MARTOS ET AL.}
\runningtitle{THE GALACTIC SPIRAL PATTERN}
\beginpage{1}
\endpage{7} 
 
\begin{document} 
    
 
\title{  
 ON THE GALACTIC SPIRAL PATTERNS: STELLAR AND GASEOUS
}

\author{MARCO MARTOS$^{1}$,
MIGUEL YA\~NEZ$^{1}$,
XAVIER HERNANDEZ$^{1}$,
EDMUNDO MORENO$^{1}$,
AND BARBARA PICHARDO$^{1}$
}

\address{$^1$Instituto de Astronom\'\i a,
Universidad Nacional Aut\'onoma de M\'exico
A. P. 70--264,  M\'exico 04510 D.F., M\'exico} 
\address{\it E-mail: marco@astroscu.unam.mx}

\abstract{
The gas response to a proposed spiral stellar pattern for our Galaxy is presented
here as calculated via 2D hydrodynamic calculations utilizing the ZEUS code in the disk plane.
The locus is that found by Drimmel (2000) from emission profiles in the K band
and at 240 $\mu m$. The self-consistency of the stellar spiral pattern was studied in
previous work (see Martos et al. 2004). It is a sensitive function of the pattern
rotation speed, $\Omega_p$, among other parameters which include
the mass in the spiral and its pitch angle. Here we further discuss the complex gaseous
response found there for plausible values of $\Omega_p$ in our Galaxy, and argue
that its  value must be close to $20~km~s^{-1}~kpc^{-1}$ from the strong self-consistency
criterion and other recent, independent studies which depend on such parameter.
However, other values of $\Omega_p$ that have been used in the literature are
explored to study the gas response to the stellar (K band) 2-armed pattern. For our
best fit values, the gaseous response to the 2-armed pattern displayed in the K band is a
four-armed pattern with complex features in the interarm regions. This response resembles
the optical arms observed in the Milky Way and other galaxies with the smooth underlying
two-armed pattern of the old stellar disk populations in our interpretation.
The complex gaseous response appears to be related to resonances in stellar orbits. Among
them, the 4:1 resonance is paramount for the axisymmetric Galactic model employed, and
the set of parameters explored. In the regime seemingly proper to our Galaxy, the spiral forcing appears
to be marginally strong in the sense that the 4:1 resonance terminates the stellar pattern, despite
its relatively low amplitude. In current work underway, the response for low values
of $\Omega_p$ tends to remove most of the rich structure found for the optimal
self-consistent model and the gaseous pattern is ring-like. For higher values than
the optimal, more features and a multi-arm structure appears.}

\maketitle

\section{INTRODUCTION}

In a recent paper (Martos et. al 2004, hereafer P1), a best fit model constrained by self-consistency
was presented for the stellar spiral pattern of our Galaxy. For the Galactic parameters
explored in that work and previous studies (Pichardo, 2003; Pichardo et. al. 2003; Pichardo,
Martos and Moreno 2004), the best fit was found for a value of the pattern speed of
$\Omega_p = 20~km~s^{-1}~kpc^{-1}$. The imposed spiral locus was the 2-armed pattern
found by Drimmel (2000). Calculating the
gas response to such pattern, we obtained a 4-armed pattern with multiple interarm features
which we identified with the optical pattern observed in our Galaxy. This
interpretation is consistent with a prediction laid down by Drimmel and Spergel (2001).
                                                                                                          
The value of the pattern speed of the Galaxy has been a matter of controversy for a long
time. From the values proposed by Lin, Yuan \& Shu (1969) of $\Omega_p = 11-13~km~s^{-1}~kpc^{-1}$,
numbers in the range of 10-60 $km~s^{-1}~kpc^{-1}$ have appeared in the literature (see,
for instance, Andrievsky et. al. 2003).
                                                                                                          
The conventional picture of the spiral pattern of the Milky Way maps at least 4 arms, named
Norma, Crux-Scutum, Carina-Sagittarius, Perseus (for a recent review see Vall\'ee 2002, who also
reports a likely pitch angle of 12 degress for this pattern). Additionaly, features such as
the Orion spur at the Solar neighborhood, have been revealed (Gorgelin \& Gorgelin 1976).
Recent data shed light into this picture, providing a deeper understanding of the Galactic
spiral structure. Drimmel (2000) presented emission profiles of the Galactic plane in the
K band and at 240 $\mu m$. The former data set, which suffers little absorption and traces density
variation in the old stellar population, is dominated by a two-armed structure with a minimum
pitch angle of 15.5 $^\circ$. At 240 $\mu m$, the pattern is consistent with the
standard four-armed model, that corresponding to the distribution of the youngest stellar
populations delineated by HII regions. 

A continued line of work by Contopolous and collaborators (see, v.g. Patsis,
Grosb{\o}l and Hiotelis 1997 and references therein) has provided a framework to study the response
of gaseous disks to spiral perturbations. In that paper, a comparison between
SPH models with Population I features observed on B images of normal, grand design galaxies,
showed that the 4:1 resonance generates a bifurcation of the arms and interarm features. Furthermore,
Contopoulos \& Grosb{\o}l (1986, 1988) had shown that the central family of periodic orbits do not
support a spiral pattern beyond the position of the 4:1 resonance, which thus determines the extent
of the pattern. Weak spirals can extend their pattern up to corotation from linear theory.
A phenomenological link betweeen resonances, the angular speed, and the stellar and gas patterns
in spirals is complemented  by the study of Grosb{\o}l \& Patsis (2001): using deep K band surface
photometry to analize spiral structure in 12 galaxies, the two-armed pattern radial extent was
found consistent with the location of the major resonances: the inner Lindblad resonance (ILR),
the 4:1 resonance, corotation and the outer Lindblad resonance (OLR). For
galaxies with a bar perturbation, the extent of the main spiral was better fitted assuming is
limited by Corotation and the OLR. 

\section{OUR MODEL}
                                                                                                          
In P1, our axisymmetric Galactic model is that of Allen \& Santill\'an (1991), which
assembles a bulge and a flattened disk proposed by Miyamoto and Nagai (1975), with a
massive spherical dark halo. The model is suitable for this particular work for its
mathematical simplicity, with closed expressions for the gravitational potential. The
main parameters are: $R_{sun} = 8.5$ kpc as the Sun's galactocentric distance; an observationally
constrained rotation curve which flattens at a moderate speed of about 200 $~km~s^{-1}$ from
a peak value of 220 $~km~s^{-1}$, and a total mass of $9\times 10^{11} M_{sun}$ within 100 kpc.                                                                                                           
We added to this mass
distribution a spiral pattern modeled as a superposition of inhomogeneous oblate spheroids
placed along different loci (for details, see Pichardo 2003; and Pichardo et al. 2003).
One locus is the fit to the K band data of Drimmel (2000), a two-armed spiral
with  a pitch angle of 15.5$^\circ$. The arms start from the tips, at a R of 3.3
kpc, and at right angles, of a line segment passing through the Galactic center. The orientation
is such that the line segment makes an angle of 20$^\circ$ with the Sun-Galactic center radius,
and is aligned with the assumed Galactic bar orientation (Freudenreich 1998; Pichardo, Martos and Moreno 2004).
The minor axis of the spheroids is perpendicular to the Galactic plane, extending up to
0.5 kpc; the major semiaxes have a length of 1 kpc. Each spheroid has a similar mass distribution,
and the central density falls along the arm locus up to a Galactic distance of
truncation of 12 kpc.
The total mass in the spiral is such that provides local ratios of spiral to background (disk)
forces of certain magnitude. Seeking reasonable values of that ratio, we used the empirical result of
Patsis, Contopoulos and Grosb{\o}l (1991); in this work, self-consistent models for 12 normal spiral
galaxies are presented, with a sample including Sa, Sb and Sc galaxies. Their Figure 15 shows
a correlation between the pitch angle of the spiral arms and the relative
radial force perturbation. The forcing, proportional to the pitch angle, is increasing
from Sa to Sc types in a linear fashion. For our pitch angle of 15.5$^\circ$, the required ratio
for self-consistency is between 5$\%$ and 10$\%$; the ratio is a function of R. The
authors consider strong spirals those in which the ratio is 6$\%$ or more. A strong spiral is
expected to terminate its pattern at the location of the 4:1 resonance for self-consistency.   
                                                                                                          
We found that, in order to obtain relative force perturbations in the 5$\%$ to 10$\%$ range, our model
requires a mass in the spiral pattern of 0.0175 $M_D$, where $M_D$ is the mass of the disk.
With that choice, our model predicts a peak
relative force of 6$\%$, and an average value, over R, of 3$\%$. Other masses were explored, but
the analysis favours this case (see below), borderline but on the weak side of the limit
separating linear (weak) and non-linear (strong) regimes considered by Contopulos \& Grosb{\o}l (1986, 1988). It is worth noticing that the latter results were obtained in galactic models quite simplified
in comparison with the one employed here. Also, the relative
amplitude of the spiral perturbation has been taken as a fixed number, a few percent of the
axisymmetric force in all work we know of for our Galaxy; for instance, Yuan (1969) proposed 5$\%$.

\section{SELF-CONSISTENCY}
                                                                                                          
Following Contopoulos \& Grosb{\o}l (1986), in Pichardo et al. (2003)
the stellar density response to the proposed spiral density distribution
was calculated assuming that trapped orbits around an unperturbated circular orbit and with the
sense of rotation of the perturber are also trapped around the corresponding central
periodic orbit in the presence of the perturbation. Thus, a series of central periodic orbits
are computed and the density response is calculated along their extension, using mass flux conservation
between two succesive orbits. Then the position of the density response maxima along each periodic
orbit, and the positions of the response maxima in the Galactic midplane, are found. These locations
are compared with the spiral locus, and the  average density response around each maximum is
calculated taking a circular radius about it of the same semiaxis of the spheroids in the model.
The ratio of the average density response and the imposed density is the merit function
for the self-consistency criterion we consider the model must satisfy.

For the spiral mass satisfying our expected spiral amplitude, the ratio of densities obtained
for $\Omega_p$ = 20 $km~s^{-1}~kpc^{-1}$ provides an excellent test of self-consistency. It
is an almost flat function of R (Pichardo 2003), with values
between 1 and 1.2 in the radial (R) interval [3.8,7.3] kpc. Large departures from this flat
response were found for other values of $\Omega_p$ close to the prefered solution. These values were
explored within  a wider range of R, [3.3, 10.3] kpc. Other
parameters (spiral mass and pitch angle, two-armed pattern) are fixed. The values were:
19, 21, 22.5, 25 $km~s^{-1}~kpc^{-1}$. In Figure 1, we show the results for the case
$\Omega_p$ = 19 $km~s^{-1}~kpc^{-1}$. On the Galactic plane, the assumed spiral pattern
and a set of stellar periodic orbits are drawn. The density maxima locations, described above, are also
pictured. The boxy-like orbit is the 4:1 resonance. In the old kinematical-wave interpretation of
orbital support for the spiral, one can see support inside the resonance, and an abrupt change
corresponding to an off-phase response to the spiral outside it. This figure may be directly
compared with Figure 1 of P1. While there are large discrepancies in the stellar response,
as the merit function for this value of $\Omega_p$ departs very much from the optimal response for
$\Omega_p$ = 20 $km~s^{-1}~kpc^{-1}$ (displayed in P1), the gas response is quite similar for both values. In other words,
self-consistency is a rather delicate test with a precise value that maximizes the merit function.
On the other hand, the  hydrodynamic gas response as shown in Figure 2 is much more robust,
varying only in a slow fashion with $\Omega_p$. This also applies to the behavior here displayed that
the pattern would be dynamically terminated at  the position of the 4:1 resonance, as
predicted by Contopulos and colaborators. Drimmel and Spergel (2001) find that the arm strength
begins to fall off inside the Solar circle at about .85 $R_{sun}$,
which, for our Galactic model-assumed $R_{sun} = 8.5$ kpc, corresponds to 7.2 kpc.

\begin{figure*}
\psfig{figure=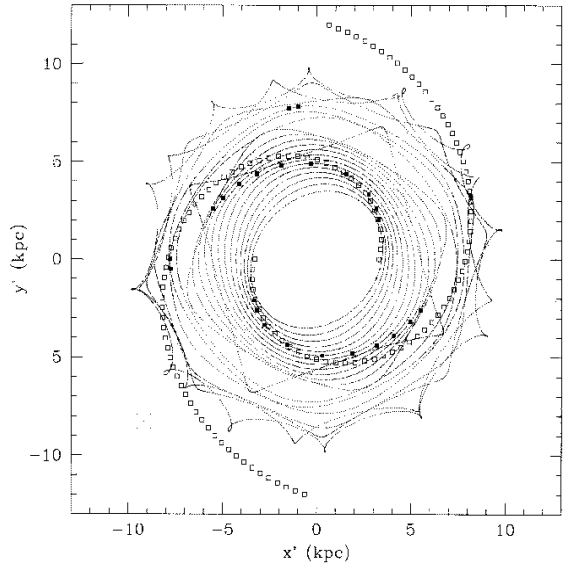,height=8cm,width=8cm}
\caption{The self-consistency analysis for $\Omega_p$ = 19 $~km~s^{-1}~kpc^{-1}$. The proposed
spiral locus is shown with open squares. A set of periodic orbits are traced with continuous
lines, and the maxima in density response are the filled (black) squares. The frame of reference
is the rotating one where the spiral pattern is at rest, with origin at the Galactic center.}
\end{figure*}

\section{THE GAS RESPONSE}

Our Figure 2 shows the gas response to the imposed pattern, whose locus is indicated
with open squares. The numerical grid is cylindrical, and covers 2 $\Pi$ radians and a radial
range of 3.3 to 15 kpc. The zone inside that inner boundary is physically meaningless,
reflecting only the initial conditions of the simulation. Calculations were performed
utilizing the 2D, hydrodynamic version of ZEUS (for a description of the code,
see Stone \& Norman, 1992a, b). There are 500 X 500 zones in the (Eulerian) grid, and
the snapshot corresponds to a time of evolution of 1.6 Gyr. The system is initialized
with outflow radial boundary conditions, and velocities from the Galactic model rotation curve,
adding the spiral source terms through a input table for ZEUS. The disk reaches a
nearly steady state rapidly as shown in this figure, which was followed  up to 3 Gyr. The
arrows are velocities in the rotating frame; the maximum size represents 212 $km~s^{-1}$.
The initial gas density is exponential, with a radial scaleheight of 15 kpc and the local
value of about 1 $cm^{-3}$. The temperature was fixed at 8000 K, and the simulation is
isothermal given the short cooling timescales compared to the dynamical timescales. The
pattern is rotating at 19 $km~s^{-1}~kpc^{-1}$.
In grey scales, the density gas response shows 4 arms and additional features.         
The shock strength along a "gaseous" arm varies along its extension.

\begin{figure*}
\psfig{figure=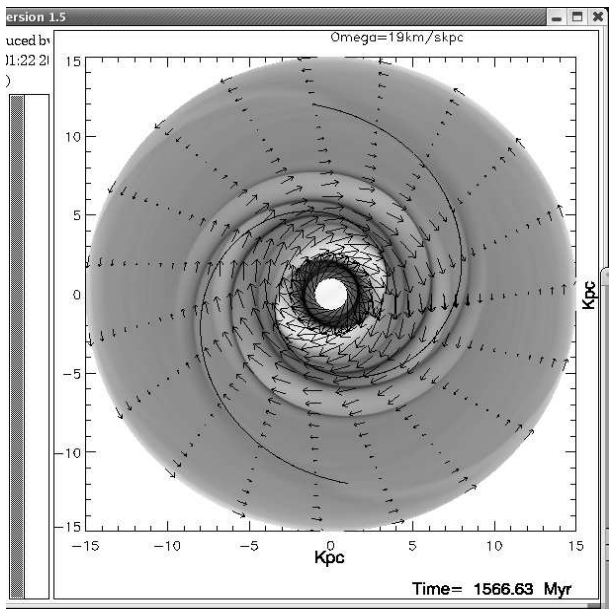,height=8cm,width=8cm}
\caption{
A simulation with the code ZEUS of the gas response to a spiral pattern with $\Omega_p$ = 19 $~km~s^{-1}~kpc^{-1}$ in the
rotating frame of the spiral pattern (the primed x-y plane of Figure 1). The arrows give the velocity field, and their size is
proportional to the speed. The maximum size in the grid corresponds to 212 $km~s^{-1}$. The
elapsed time is 1.6 Gyr. In grey scales, dense zones correspond to dark regions. This
simulation is isothermal with 500 X 500 zones in cylindrical coordinates. The imposed spiral pattern
is illustrated with the solid line.
 }
\end{figure*}

In work currently underway, we have found (Y\'a\~nez and Martos 2004) that for low
$\Omega_p$ (of the order 10 $km~s^{-1}~kpc^{-1}$), much of the rich structure is
removed, rendering a ring-like pattern. High values (of the order of 40 $km~s^{-1}~kpc^{-1}$)
tend to produce a more complex gas pattern.
This is seemingly revealing the strong link between resonances in the stellar orbits
and the gas structure.
We notice that the strength of the shocks will considerably diminish in a full 3D, MHD simulation
(the inclusion of the vertical direction and magnetic field was considered in Martos \& Cox 1998).
Secondly, the response is just too sensitive to the details of the potential
to be conclusive. Recent  simulations in realistic Galactic models are
scarse: G\'omez-Reyes \& Cox (2002) employed the Galactic model of Dehnen and Binney (1998).
However, their value of $\Omega_p$ (12 $km~s^{-1}~kpc^{-1}$) places the 4:1 resonance
beyond 22 kpc, far out from the observed pattern extent.
Another important
component not included yet in these simulations is the Galactic bar, which we recently modeled in three ways
(Pichardo, Martos \& Moreno 2004). As discussed in P1, we do not expect a strong effect at the
Solar circle, but the bar will affect the dynamics not too far away from that position towards
the inner Galaxy. A plausible assumption is that R = 3.3 kpc corresponds to the bar corotation.
If the study of  Grosb{\o}l \& Patsis (2001) applies, one would say
that the termination of the ``main" spiral pattern with a bar probably corresponds to
the position of corotation or the OLR.
In the following, we discuss independent studies that are consistent
with the picture here described.

\section{STAR FORMATION IN THE GALAXY}
                                                                                                          
For values of $\Omega_p$ = $20~km~s^{-1}~kpc^{-1}$ or closer, one can obtain from
Figure 2 the surface density along a circle with
the radius of the Sun orbit. In the most simplistic approximation (the orbit has
radial excursions of the order of 2 kpc), there are two main peaks of similar densities,
and several local maxima. The mass contrast is consistent with K-band observations (Kranz, Slyz \& Rix 2001),
who quote an arm/interarm density contrast for the old stellar population of 1.8 to
3 for a sample of spiral galaxies. Interestingly, these two peaks have surface densities in
agreement with the expected threshold density for star
formation, with a value of approximately 10 M$\odot~pc^{-2}$ (Kennicutt 1989),
in which we are considering the reduction in the shock compression due to the magnetic field
and the 3D dynamical effects. Other local maxima are factors of 3 or more lesser than that value,
making any burst of star formation a less likely event.
Now, with a $\Omega_p$ = $20~km~s^{-1}~kpc^{-1}$, the time baseline for
a circular orbit of that radius is very approximately 1 Gyr,
with our assumed Solar galactocentric distance. We found then two similarly spaced episodes of vigorous star
formation over the last Gyr. To test this prediction, we compare with a recent star formation
history of the Hipparcos Solar neighbourhood, by Hern\'andez, Valls-Gabaud and Gilmore (2000).
Using Bayesian analyisis techniques to derive the star formation history over the last
3 Gyr, they find an oscillatory component of period of nearly 0.5 Gyr, in remarkable agreement
with our result, over a small level of constant star formation.
In consistency with these two independent determinations, de la Fuente Marcos and de la Fuente
Marcos (2004) obtain, by studying the age distribution of young globular clusters, a periodicity
in the recent star formation history at the Solar circle of 0.4  $\pm0.1$ Gyr.

\section{COSMIC RAY DIFFUSION FROM SPIRAL ARMS}
                                                                                                          
Shaviv (2002) finds that the CR flux reaching our Solar System should periodically increase
with each crossing of a Galactic spiral arm. Along the same lines of last section,
over a time baseline of the past 1 Gyr, we
added our estimated magnetic field compression form Martos \& Cox (1998) to plot the expected
synchrotron flux variations moving along the Solar circle and assuming the mass distribution fixed in time.
There are 6 local maxima in our plot with fluxes that are higher than today's flux.
This is the same number of peaks satisfying that condition in
Shaviv's work, who plots the ratio cosmic ray flux/today's cosmic ray flux obtained from a
sample of 42 meteorites, which is related there to climate changes in Earth.
Shaviv (2002) reports a period of 143 Myr for the episodes (crossings) from meteorites data.
In our framework, crossings occur on the average every 167 Myr, not equally spaced in time.
Interestingly, there is a delay of about 25 Myr to be expected in synchrotron emission because of the time
after the ionizing photons are emitted, plus the time for cosmic ray diffusion, as discussed
in Shaviv (2002).
Shaviv assumes 4 arms and a termination of the pattern at the 4:1 resonance. This is
the behavior we found for the stellar self-consistency, but from the fact that the K band
and the optical pattern are observed to extend further out, it is likely that the
analysis must be revised.

\section{CONCLUSIONS}
                                                                                                          
To clarify matters, a higher difference in speeds between the pattern and the local speed of
the large scale flow of gas, which increases from corotation inwards, does not necessarily 
imply a stronger shock. What is critical (see Martos \& Cox 1998) is the
comparison between the kinetic energy of the gas flow and the potential well the spiral
arm represents at some locations. We stress the point that, with the addition of
the spiral pattern modeled as a structure from a mass distribution, vs the usual perturbing
term commonly used in the literature, the resulting gravitational potential becomes a
strong function of the position in the disk, even with the smooth, continuous distribution
we propose. Reality must be more complex. As a consequence, an arm does not always represent
a potential well, and along a given spiral arm, the compression from the gas flow encounters
will be a sensitive function of the coordinate along the locus, specially at R in the
range of 4 or more kpc. This is reflected in Figure 2.
                                                                                                          
With uncertainties in structural parameters in our Galaxy, one can hardly expect exact
agreement between models. Our very limited modeling effort points to a particular value
of $\Omega_p$, 20 $~km~s^{-1}~kpc^{-1}$ that maximizes the self-consistency criterion we utilized.
As found by Contopoulos \& Grosb{\o}l (1988), that self-consistency analysis is improved
introducing dispersion of velocities; this is a realistic effect that can be neglected arguing
that in strong spirals nonlinearity dominates. For our Galaxy, we did not explore the region
betweeen the 4:1 resonance and corotation in enough detail to claim that our Galaxy is a strong spiral.
It appears rather weak from the observed patterns.
However, the response at the best $\Omega_p$ is so flat that does not seem to need the aid of
a dispersion of velocities. Other values of $\Omega_p$ could be improved by this effect and
give as good a response. In particular, from our results, values close to             
20 $~km~s^{-1}~kpc^{-1}$ appear as reasonable candidates, as that of Figure 1.
Much more work remains to be done. The inclusion of the bar should provide
further insight into the structural details, but the reported, large differences between
the angular speeds of bar and spiral complicate the dynamical coupling (Bissantz, Englemaier
\& Gerhard 2003 find a best model coupling at speeds of 60 $~km~s^{-1}~kpc^{-1}$ for the
bar and 20 $~km~s^{-1}~kpc^{-1}$ for the spiral, using SPH models).
                                                                                                          
The detailed gaseous response for different values of $\Omega_p$, in the presence of
large scale magnetic fields is work currently underway and will be presented elsewhere
(Y\'a\~nez and Martos 2004). The link betweeen resonances in the stellar orbits and
the complex gas response is also worth exploring further, as the study of Chakrabarti,
Laughlin and Shu (2003) indicates.

\acknowledgements{
E. Moreno, M. Martos, B. Pichardo, M. Y\'a\~nez acknowledge financial support from
{\bf CoNaCyT} grant {\tt 36566-E}, and UNAM-DGAPA grant IN114001. M. Martos
thanks M. Norman, D. Clarke and associated group for advise in the use of ZEUS for this
particular application, and helpful discussions with P. Grosb{\o}l, P. Patsis, E. Athanassoula,
D. Cox, G. G\'omez-Reyes, L. Sparke, F. Masset and S. Shore.}

\end{document}